\documentclass[traditabstract,printer]{aa}
\usepackage{txfonts}
\usepackage{graphicx}
\usepackage{color}
\usepackage{natbib}

\begin{document}

\title{Spectroscopic study of the elusive globular cluster ESO452-SC11 and its surroundings}

\author{
Andreas Koch\inst{1} 
  \and Camilla Juul Hansen\inst{2}
  \and Andrea Kunder\inst{3,4}
  }
  
\authorrunning{A. Koch, C.J. Hansen, \& A. Kunder}
\titlerunning{Spectroscopy of ESO452-SC11}
\offprints{A. Koch;  \email{a.koch1@lancaster.ac.uk}}
\institute{Department of Physics, Lancaster University, LA1 4YB, Lancaster, UK
  \and Dark Cosmology Centre, Niels Bohr Institute, University of Copenhagen, Juliane Maries Vej 30, DK-2100, Copenhagen 
  \and Leibniz-Institut f\"ur Astrophysik Potsdam, An der Sternwarte 16, 14482 Potsdam, Germany   
  \and Saint Martin's University, Old Main, 5000 Abbey Way SE, Lacey, WA 98503, USA
   }
\date{Received 21 March 2017 / Accepted 24 April 2017}
\abstract {Globular clusters (GCs) have long been recognized as being amongst the oldest objects in the Galaxy. 
As such, they have the potential of playing a pivotal role in deciphering the Milky Way's early history.
Here we present the first spectroscopic study of the low-mass system ESO452-SC11 using the AAOmega multifibre spectrograph at medium resolution. 
Given the stellar sparsity of this object and the high degree of foreground contamination due to its location toward the Galactic bulge,  very few details are known for this cluster --  there is no consensus, for instance, about its age, metallicity, or its association with the disk or bulge.
We identify five member candidates based on common radial velocity, calcium-triplet metallicity, and position within the GC.
Using spectral synthesis, the measurement of accurate Fe-abundances from Fe-lines, and 
abundances of several $\alpha$-, Fe-peak, and neutron-capture elements (Si, Ca, Ti,Cr, Co, Ni, Sr, and Eu) is carried out, 
albeit with large uncertainties. 
We find that two of the five cluster candidates are likely non-members, as they have deviating iron abundances and [$\alpha$/Fe] ratios.  The cluster mean heliocentric velocity is 19$\pm$2 km\,s$^{-1}$ with a velocity dispersion of 
2.8$\pm$3.4 km\,s$^{-1}$, a low value in line with its sparse nature and low mass. 
The mean Fe-abundance from spectral fitting is $-0.88\pm0.03$ dex, where the spread is driven by observational errors.
Furthermore, the $\alpha$-elements of the GC candidates are marginally lower than expected 
for the bulge at similar metallicities. 
As spectra of hundreds of stars were collected in a 2-degree field centered on ESO452-SC11, 
a detailed abundance study of the surrounding field was also enabled.  The majority of the non-members 
have slightly higher [$\alpha$/Fe] ratios, in line with the typical nearby bulge population. 
A subset of the spectra with measured Fe-peak  abundance ratios shows a  large scatter 
around solar values, albeit with large uncertainties. Furthermore, our study provides the first systematic measurements 
of strontium abundances in a Galactic bulge GC. 
Here, the Eu and Sr abundances of the GC candidates 
are broadly consistent with a disk or bulge association. 
Recent proper motions and our orbital calculations place ESO452 on an elliptical orbit in the central 3 kpc of the Milky Way, establishing a firm 
connection with the bulge. 
 Finally, while the radial velocities and preferential position of a dozen of stars outside the GC radius appear to imply the presence of extra-tidal stars, their significantly different 
chemical composition refutes this hypothesis.}
\maketitle 
%
%
%
%%%%%%%%%%%%%%%%%%%%%%%%%%%%%%%%%%%%%%%%%%%%%%%%%%%%%%%%%%%%%%%%%%%%%%%%%%%%%
%
%
%
\section{Introduction}
Star clusters, in particular the old globular clusters (GCs), are important probes of the formation and structure of our Galaxy. 
Especially when studying objects toward 
 the Milky Way central regions, the interface of various Galactic components becomes important, 
where the bulge,  thin and thick disks, and the halo tend to overlap \citep[e.g.,][]{ChibaBeers2000,Koch2008,Ness2013}. 
While it is difficult to distinguish these structures purely based on their metallicity or kinematics, chemical abundance tagging 
provides a powerful tool to study the evolution and distinction of the individual components in detail \citep[e.g.,][]{NissenSchuster2010,McWilliam2016}.

ESO452-SC11 (in the following ESO452 for brevity) is a poorly
studied object located at $\sim$2 kpc from the 
Galactic Center at low latitude  that was discovered in the ESO/Uppsala B survey of \citet{Lauberts1981}. 
Table~1 lists the fundamental properties of this GC. 
\begin{table}[htb]
\caption{Properties of  ESO452-SC11 from the literature and  this work.}             
\centering          
\begin{tabular}{ccc}     
\hline\hline       
 Parameter & Value & Reference \\
 \hline
( $\alpha$, $\delta$) (J2000.0) & (16:39:25.45,$-$28:23:55.3) & (1) \\
($l$,$b$) &  (351.91$\degr$,+12.10$\degr$)  & (1) \\
R$_{\rm GC}$, R$_{\odot}$ & $\sim$2 kpc , $\sim$7 kpc  & (1),(2) \\
 r$_c$,  r$_h$, r$_t$   & 0.5$\arcmin$, 0.5$\arcmin$, 5.0$\arcmin$  & (1)\\
M$_V$ & $-4.02$ mag     & (1) \\
E(B$-$V) & 0.521 mag, 0.448  mag & (3), (4) \\
$<$v$_{\rm HC}$$>$, $\sigma$ &  19$\pm$2 km\,s$^{-1}$, 2.8$\pm$3.4 km\,s$^{-1}$ & (5) \\
$<$[Fe/H]$>_{\rm (SP\_ACE)}$ & $-0.88$ dex & (5)\\
\hline
\end{tabular}
\tablefoot{References: (1) \citet[][2010 version]{Harris1996}; (2) \citet{Cornish2006}; (3) \citet{Schlegel1998}; (4) \citet{Schlafly2011}; (5) This work.}
\end{table}
The recent study of \citet{Cornish2006} found that  depending on the exact isochrones used, the age of this GC was still
difficult to constrain, leaving a possible range between 9 and 16 Gyr. Similarly, a  metallicity between
$-1.4$ and $-0.4$ dex and heliocentric distances between 6.6 and 7.5 kpc were in the permitted parameter space of this sparse object's 
color-magnitude diagram (CMD). Any conclusions on its properties were also sensitive to the $\alpha$-enhancement of the isochrones 
in the fitting of \citet{Cornish2006}. 

Within this broad parameter range, ESO452 was placed within the bulge-halo transition, at the lower end of the bulge metallicity distribution function (MDF), 
at a similar age to ``somewhat young halo clusters'' \citep{Cornish2006}. 
It has therefore been suggested 
that this unique GC could have been placed within the bulge by an interaction with a satellite galaxy \citep[see also][]{Marin-Franch2009,Ferraro2009}, 
which still needs to be consolidated by a chemodynamical study of ESO452.

Here, we report on our large-field-of-view (2$\degr$), multi-object spectroscopic effort with the AAOmega spectrograph
to capture this elusive object.
Our main focus thus lies on establishing membership with the GC and on 
kinematically and chemically characterizing it, also in the context of the surrounding field. 
This paper is organized as follows: in Sect. 2 we introduce our spectroscopic data set, and in Sect. 3 we use kinematic and metallicity information to assess stars that are probable members of the GC. 
Section 4 is dedicated to a stellar parameter and abundance analysis of the cluster stars and the foreground component, while in Sects.
5 and 6 we 
investigate possible correlations of our abundance results with kinematics and test for the presence of extra-tidal stars. 
In Section 7 we conclude.
\section{Observations and data reduction}
Prompted by the large field of view of our pointings, we opted to select targets from the GC (J$-$K$_S$,K$_S$) CMD (see Fig.~1) 
that we assembled from the Two Micron All Sky Survey \citep[2MASS;][]{Cutri2003}. 
\begin{figure}[htb]
\centering
\includegraphics[width=0.7\hsize, clip=true]{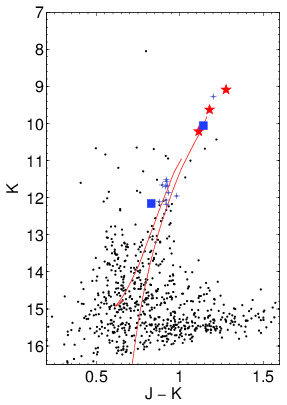}
\caption{2MASS CMD of ESO452-SC11 within its tidal radius. Targets within this radius are shown as blue  crosses, {red symbols 
denote our most central bona fide member candidates, and the blue squares are two additional candidates with similar kinematics but different chemistry. 
Also shown is a Dartmouth isochrone \citep{Dotter2008} with the cluster parameters from the literature and as found in this work.}}
\end{figure}
{Selection boxes were drawn encompassing the cluster red giant branch (RGB) from the tip RGB
to $\sim$3 mag fainter, thus  approximately one magnitude above its red clump. Furthermore,  priority was given to stars with small radial distances to the nominal GC center so as 
to maximize the number of member stars. This is only marginally efficient} because of the overall small tidal radius of r$_t=5\arcmin$ 
of ESO452 (Table~1)  and the sparse nature of this object.
Furthermore, at the low Galactic latitude of +12$\degr,$  
the yield of  this procedure is still hampered by a large contamination 
with Galactic foreground stars. 

All our data were taken on June 08, 2013, with the AAOmega multifibre spectrograph at the 3.9 m  Anglo-Australian Telescope (Siding Spring, Australia). 
A single mask  was created for the Two Degree Field (2dF) fibre positioner,
containing 363 science fibres and 25 sky positions. 
The dual setup we used employed the blue 3200B grating, centered at 4100 \AA,~and 
the red 1700D grating, centered at 8600 \AA,~so as to comprise the neutron-capture elements Sr and Eu in the blue \citep{CJHansen2015} and the prominent calcium triplet (CaT) lines in the red. 
The  exposure time was chosen as 4$\times$20 minutes. 
The data were reduced using the 2dfdr pipeline provided by the AAO, which performs standard reduction steps from quartz-flatfielding and wavelength 
calibration via arc-lamp exposures, sky subtraction using the dedicated sky fibres, to optimal extraction of the science spectra. 
The final spectra cover the wavelength ranges of 3970--4250 and 8350--8800 \AA, respectively, with slight variations depending on the exact location of the spectra on the CCD. 
The resolving power of the blue and red spectra is $\sim$9000 and 10600, as measured from the width of the calibration arc lamps. 
Finally, the median signal-to-noise ratio (S/N) of our final extracted spectra is {15 px$^{-1}$} in the blue and 50 px$^{-1}$ on the red chip, respectively. 
While the blue spectra reach S/Ns up to 20 px$^{-1}$, 
 the values on the red CCDs  reach as high as 170 px$^{-1}$. 
\section{Radial velocities, metallicities, and membership}
In the following, we assess the membership, metallicity,  and basic kinematic properties that can be derived from our 
spectra. 
\subsection{Radial velocity}
Individual heliocentric radial velocities of the target stars were determined via cross-correlation
of the three prominent CaT lines against a Gaussian template, using the IRAF {\em fxcor} task \citep[e.g.,][]{Kleyna2004}. 
The typical median velocity error achieved in this way is 1.4 km\,s$^{-1}$. Measurements of individual stars are listed in Table~2,
and Fig.~2 shows the run of the heliocentric radial velocities with radial distance from the center of ESO452.
\begin{table*}[htb]
\begin{center}          
\caption{Properties of the target stars.}             
\begin{tabular}{rccccc}     
\hline\hline       
& ($\alpha$ , $\delta$) & $r$\tablefootmark{b} &  v$_{\rm HC}$ & $\Sigma$EW (CaT) & [Fe/H]$_{\rm CaT}$  \\
 \raisebox{1.5ex}[-1.5ex]{Star\tablefootmark{a}} &
  (J2000.0) &  [$\arcmin$] & [km\,s$^{-1}$] & [m\AA] & [dex] \\ 
\hline
\multicolumn{5}{c}{Member candidates}\\
\hline
 55 & 16:39:27.9 $-$28:24:07.1 & 0.34 & 20.0$\pm$1.4 & 6.19$\pm$0.09 & $-$1.15$\pm$0.20 \\ 
 56 & 16:39:38.9 $-$28:26:20.8 & 2.84 &  14.6$\pm$1.2 & 5.74$\pm$0.21 & $-$1.01$\pm$0.22 \\ 
204 & 16:39:24.1 $-$28:23:07.6 & 0.81 &  17.7$\pm$1.3 & 6.29$\pm$0.11 & $-$1.22$\pm$0.20 \\ 
362 & 16:39:25.3 $-$28:23:38.5 & 0.28 &  20.5$\pm$1.5 & 6.45$\pm$0.91 & $-$1.24$\pm$0.44 \\ 
396 & 16:39:44.3 $-$28:25:10.9 & 2.42 &  14.2$\pm$1.8 & 6.34$\pm$0.13 & $-$1.11$\pm$0.21 \\ 
\hline
\hline
\multicolumn{5}{c}{Fore- and background candidates}\\
\hline
2 & 16:41:47.8 $-$28:25:02.2 & 15.7 &  $-$12.2$\pm$1.6 & 5.56$\pm$0.14 & \dots \\ 
\hline
\end{tabular}
\tablefoot{
Table 2 is available in its entirety in electronic form at the CDS.
\tablefoottext{a}{IDs corresponding to our fibre numbers.}
\tablefoottext{b}{Radial distance from the nominal cluster center.}
}
\end{center}
\end{table*}
\begin{figure}[htb]
\centering
\includegraphics[width=1\hsize, trim= 0 0 0 0, clip=true]{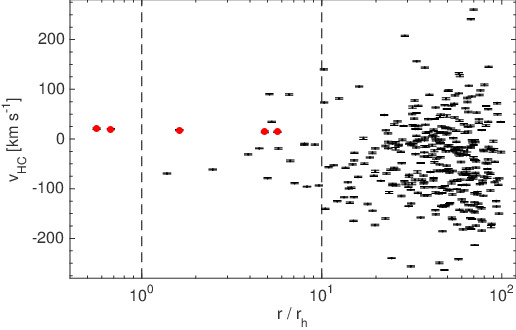}
\caption{Heliocentric radial velocities versus radial distance from the cluster center in units of its 
half-light radius. The dashed lines illustrate the GC half-light  and nominal 
tidal radius. Our five {\em \textup{bona fide}} candidates are highlighted with red circles.}
\end{figure}

Guided by the very similar velocities of the two innermost stars, well within the half-light radius, we identify 
five stars within the tidal radius  at similar velocity with individual values that lie between 14.2 and  20.5 km\,s$^{-1}$. 
All other stars within this GC radius deviate by more 
than 15 km\,s$^{-1}$ from these candidates, so that we define these five objects as the cluster population based on 
their kinematics and position.
The mean systemic velocity of the five candidate members is 
17.5$\pm$1.1 km\,s$^{-1}$ and their radial velocity dispersion  is low, 
at 2.1$\pm$1.0 km\,s$^{-1}$, which is fully in line with the low mass of the cluster \citep{Pryor1993,Dubath1997}.

In Section~4 we show that it is advisable to cull two candidates based on their [Fe/H] and [$\alpha$/Fe] ratios because these
are significantly different from the remainder of the GC sample.  Retaining only the three best candidates, 
a mean velocity of 19.4$\pm$1.8 km\,s$^{-1}$ is found with a dispersion of 2.8$\pm$3.4 km\,s$^{-1}$. 
As these three stars are those that are located closest to the cluster center, while the two most deviant stars lie at 2.5$\arcmin$ -- five half-light radii away from the GC center -- it is fair to adopt these three stars as the most probable cluster members.
\subsection{Calcium triplet metallicity}
The prominent near-infrared CaT lines at $\sim$8500 \AA~are powerful  
indicators of the global metallicity in red giants \citep[e.g.,][]{Armandroff1988}. Generally starting from 
some linear combinations of the line equivalent widths  (EWs),   
the stars' luminosity also needs to be accounted for in order to remove higher order dependencies on 
surface gravity. While this is usually done in terms of the V-band magnitude above the horizontal branch (V$-$V$_{\rm HB}$),
 our work employed infrared colors throughout, making this standard optical  approach poorly calibrated. 
Infrared (2MASS) K$_S$-magnitudes instead of  V- or I-data  have more advantages, since K$_S$ is less sensitive to reddening;  
thus, any  dependence of the reduced width, $W'$, on intracluster reddening is reduced \citep{Warren2009}. 

Various calibrations of the metallicity scale based on infrared magnitudes have been proposed  \citep[e.g.,][]{Warren2009,Mauro2014,Vasquez2015}.
 Here, we chose to adopt the study of \citet{Vasquez2015}, which is based on 
Galactic bulge stars, as it covers a broad range in metallicity from $-2.3$ to 0.7 dex. 
These calibrations to the metallicity scale of \citet{Carretta2009} read
\begin{displaymath}
{\rm [Fe/H]}\,=\,-3.15+0.432W' + 0.006W'^2, 
\end{displaymath}
 where the reduced width $W'=\Sigma\,{\rm EW} + 0.384\,(K\,-\,K_{\rm RC})$ 
and the CaT line strength was defined as the straight sum of the two strongest lines, $\Sigma {\rm EW} =  EW_{8542} + EW_{8662}$. 
Errors on EW measurements and photometry were propagated through this formalism and added in quadrature to 
the R.M.S. scatter of 0.19 dex in the calibrations, as evaluated by \citet{Vasquez2015}. 

One important issue pertains to the magnitude of the red clump, K$_{\rm RC}$. 
Our 2MASS CMD and a suitable set of old moderately metal-poor 
{isochrones \citep[$\sim -1$ dex][]{Dotter2008} show
 a sparse red clump at K$_S$=15 mag, which we adopt as the best value in our subsequent metallicity calibration.
However, the optical data of 
\citet{Cornish2006} imply distance moduli that vary by 0.6 mag depending on the metallicity and 
$\alpha$-enhancement used in their isochrone fitting. 
Here, we account for such uncertainties by also testing the CaT calibration with  a fainter red clump level of K$_S$=15.6 mag. 
This would systematically lower the resulting metallicity scale by 0.11 dex.} 

The resulting mean metallicity from the CaT of the five member candidates is $-1.15$ dex with a 1$\sigma$ scatter of 0.09 dex. 
As above, we consider the three most central stars as best members, and obtain a mean CaT metallicity of $-1.19$ dex (scatter of 0.05 dex) 
for them. 
Comparison with the median measurement error of 0.20 dex indicates that this scatter in the metallicities is most likely 
only due to the uncertainty in our CaT determination and is not an intrinsic spread, 
in line with the very low overall mass of the system \citep[e.g.,][]{Koch2012}.
We did not attempt to assign any metallicities from the CaT to the Galactic contaminants, since the above 
calibrations are only valid at the assumed red clump magnitude, thus the distance of the GC. The Galactic stars, in turn, 
are located at random unknown distances to the observer. 

The CMD study of \citet{Cornish2006} attempted isochrone fits of ESO452, but the sparse
nature of this  object left room for a wide range of possible matches, with the cluster metallicity reaching from 
$-1.4$  up to $-0.4$ dex. Our CaT metallicity of $-1.15$ dex  places ESO452 well within  
this parameter space.  This metallicity is typical of globular clusters in the bulge (see also Sect.~5) as
well as the metal-weak thick disk. In the next sections we further refine our measurement based on spectral synthesis. 
\section{Stellar parameters and chemical abundances}
The near-infrared region around the CaT at intermediate resolution contains 
a wealth of spectral information in the form of several temperature- and gravity-sensitive features and 
chemical abundance tracers \citep[][]{Zwitter2004,Shetrone2009,Ruchti2010,Hendricks2014}. 
Thus we also attempted to extract stellar parameters and chemical abundance ratios of several elements from the spectra 
in addition to the CaT metallicities we have obtained above. 
\subsection{SP\_ACE}
To this end, we first used the 
SP\_ACE code \citep{Boeche2016}, which was originally designed to complement the 
Radial Velocity Experiment (RAVE) pipeline \citep{Siebert2011,Kunder2017}.  RAVE operates 
in a similar wavelength region at slightly lower resolution
than our data, but SP\_ACE can been adapted to a much broader range of applicabilities. 

In brief, EWs of each line in the input spectral region are  first fit by a polynomial  as a function of the stellar parameters
T$_{\rm eff}$, log\,$g$, [M/H], and [X/Fe] for chemical elements X. 
The actual SP\_ACE routine then constructs curves of growth as a function of these  parameters and determines the best fit in a $\chi^2$-sense. 
As detailed in \citet{Boeche2016}, the microturbulence is computed by SP\_ACE {\em \textup{ad hoc}} as a function of temperature and gravity 
\citep[see also][]{Koch2009,Jofre2014}.

In practice, we restricted the spectral range to be fitted to 8505--8535, 8547--8660, and 8670--8702 \AA, thereby avoiding the strong CaT lines because of their very strong wings, which can cause difficulties in precise determinations of the surface gravity and/or the Ca-abundance 
of the stars. 
{
Furthermore, we thus avoided the TiO$\varepsilon$ band at 8432 \AA~that can affect abundance analyses of cold metal-rich stars.
To quantify the overall magnitude of its influence, we computed a TiO$\varepsilon$ index following \citet{Sharples1990} and \citet{Kunder2012}.
As a result, this index is scattered around zero, which shows that the pseudo-continuum in this spectral range is predominantly flat and thus 
not affected by TiO absorption, as is expected because most our stars are warmer than $\sim$3800 K \citep[see, e.g., Fig. 6 in][]{Kunder2012} .  
}

In the employed fitting region, up to eight Fe\,{\sc i} plus {\sc ii}, nine Si\,{\sc i}, 4 Ca\,{\sc i}, nine Ti\,{\sc i}, 3 Cr\,{\sc i}, three Co\,{\sc i}, and three Ni\,{\sc i} 
features could be measured depending on the stellar type and the S/N of the spectra.
Figure~3 shows two typical stars of our sample and the respective best-fit synthetic spectra.
\begin{figure}[htb]
\centering
\includegraphics[width=1\hsize,clip=true]{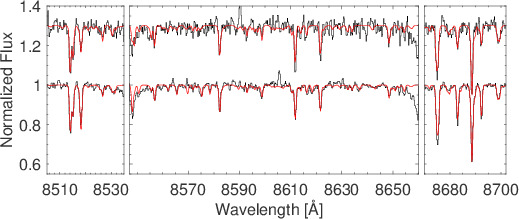}
\caption{Observed sample spectra (black) with the best-fit spectra from SP\_ACE overlaid in red. Note the broken wavelength axis.}
\end{figure}

All the stellar parameters of the member fore- or background stars thus derived are
listed in Table~3, and the chemical abundance ratios are given in Tables~4 and 5.
Owing to the statistical method of abundance fitting in SP\_ACE, the code returns asymmetric error bars, which we list in the 
tables as  lower and upper bounds. Finally, the number of lines, N,  detected in each spectrum is also tabulated. 
\begin{table*}
\caption{Stellar parameters from our photometric analysis and from SP\_ACE for the member candidates in ESO4522-SC11.}
\begin{center}
\begin{tabular}{ccclccccc}
\hline
\hline
Star & J$-$K & \multicolumn{2}{c}{T$_{\rm eff} [K]$} && log\,$g$ & $\xi$  & S/N & $\chi^2$ \\
\cline{3-4}
&  & Alonso  & SP\_ACE &&   & [km\,s$^{-1}$] & & \\
\hline
\hline
\phantom{3}55  & 1.114 & 3982 & 3983$_{-83}^{+41}$   &&  1.64$_{-0.47}^{+0.11}$ & 1.47 & 86 & 1.13 \\
\phantom{3}56  & 0.829 & 4619 & 4794$_{-116}^{+148}$ &&  3.30$_{-0.28}^{+0.37}$ & 1.24 &  37 & 1.05 \\
          204  & 1.179 & 3870 & 3928$_{-43}^{+90}$   &&  1.04$_{-0.13}^{+0.51}$ & 1.47 & \llap{1}10 & 1.10 \\
          362  & 1.278 & \ldots & 3880$_{-31}^{+80}$   &&  1.11$_{-0.17}^{+0.36}$ & 1.42 & 91 & 1.03 \\
          396  & 1.141 & 3910 & 3978$_{-63}^{+61}$   &&  1.63$_{-0.43}^{+0.13}$ & 1.47 & 84 & 1.00 \\
\hline
\end{tabular}
\end{center}
\tablefoot{
Table 3 is available in its entirety in electronic form at the CDS, including
  the fore- and background stars.}
\end{table*}
\begin{table*}
\caption{Stellar abundances from SP\_ACE for the member candidates in ESO4522-SC11.}
\begin{center}
\begin{tabular}{ccccccccccccc}
\hline
\hline
Star & 
[Fe/H]  & $\sigma_{{\rm Fe}}$ & N$_{\rm Fe}$ & 
[Si/Fe] & $\sigma_{{\rm Si}}$ & N$_{\rm Si}$ & 
[Ca/Fe] & $\sigma_{{\rm Ca}}$ & N$_{\rm Ca}$ & 
[Ti/Fe] & $\sigma_{{\rm Ti}}$ & N$_{\rm Ti}$ \\
\hline
\hline
\phantom{3}55 &            $-$0.96 & $\pm$0.05 & 3 &      0.75 & $\pm$0.10         & 3 &  0.37 & $^{+0.35}_{-0.20}$ & 4 &           0.03 & $^{+0.27}_{-0.14}$ & 1\\
\phantom{3}56 &  \phantom{$-$}0.15 & $\pm$0.08 & 4 & \llap{$-$}0.03 & $\pm$0.14     & 1 &  0.18 & $^{+0.28}_{-0.23}$ & 4 & \llap{$-$}0.04 &         $\pm$0.24  & 1\\
204           &            $-$0.85 & $\pm$0.05 & 4 &      0.44 & $\pm$0.15         & 7 &  0.11 & $^{+0.38}_{-0.27}$ & 4 &           0.05 &         $\pm$0.10  & 1\\
362           &            $-$0.82 & $\pm$0.05 & 4 &      0.39 & $\pm$0.16         & 5 &  0.14 & $^{+0.38}_{-0.23}$ & 4 &           0.11 & $^{+0.06}_{-0.23}$ & 1\\
396           &            $-$0.52 & $\pm$0.05 & 4 &      0.46 & $^{+0.21}_{-0.13}$ & 6 &  0.12 & $^{+0.31}_{-0.18}$ & 4 &          0.04 & $^{+0.09}_{-0.20}$ & 1\\
\hline
\end{tabular}
\end{center}
\tablefoot{
Table~4 is available in its entirety in electronic form at the CDS, including
  the fore- and background stars.  }
\end{table*}
\begin{table*}
\caption{Same as Table~4, but for the elements Cr, Co, and Ni from SP\_ACE. The Sr and Eu abundances, in turn, are based on synthetic fits.}
\begin{center}
\begin{tabular}{cccccccccccccc}
\hline
\hline
Star & 
[Cr/Fe] & $\sigma_{{\rm Cr}}$ & N$_{\rm Cr}$ & 
[Co/Fe] & $\sigma_{{\rm Co}}$ & N$_{\rm Co}$ & 
[Ni/Fe] & $\sigma_{{\rm Ni}}$ & N$_{\rm Ni}$ &
[Sr/Fe] & $\sigma_{{\rm Sr}}$ &  [Eu/Fe] & $\sigma_{{\rm Eu}}$ \\ 
\hline
\hline
\phantom{3}55 &  \phantom{$-$}0.00 & $^{+0.45}_{-0.25}$ & 3 &         \ldots &    \ldots & 0 &                  \ldots &             \ldots & 0 &   0.20 &   0.20 &   \ldots & \ldots \\
\phantom{3}56 &            $-$0.40 &          $\pm$0.50 & 3 & \llap{$-$}0.20 & $\pm$0.55 & 2 &                 $-$0.35 &          $\pm$0.80 & 2 & \ldots & \ldots &   \ldots & \ldots \\
204           &            $-$0.18 &          $\pm$0.23 & 3 &            0.31 & $\pm$0.25 & 3 &           $-$0.41 &          $\pm$0.56 & 2 &   0.00 &   0.09 &     0.40 &   0.14 \\
362           &            $-$0.20 &          $\pm$0.23 & 3 &            0.45 & $\pm$0.23 & 2 &           $-$0.20 &          $\pm$0.51 & 2 & \ldots & \ldots &  \ldots  & \ldots \\
396           &            $-$0.30 &          $\pm$0.26 & 3 &         \ldots &    \ldots & 0 & \phantom{$-$}0.16 & $^{+0.60}_{-0.33}$ & 2 &   0.10 &   0.30 & $>-$0.30 & \ldots \\
\hline                                                              
\end{tabular}
\end{center}
\tablefoot{
Table~5 is available in its entirety in electronic form at the CDS, including
  the fore- and background stars.  }
\end{table*}
\subsection{T$_{\rm eff}$ and log\,$g$ from photometry}
To test the robustness of the parameters obtained from SP\_ACE, we attempted two additional independent methods to estimate the temperatures and gravities of the all the observed stars. The short spectral
ranges from the two AAOmega settings make it hard to determine  [Fe/H] accurately, and therefore using the $J-K$ color-T calibration is advantageous owing to its low metallicity dependence. 

The temperatures were solely based on 2MASS photometry, since we only have infrared J, H, and K magnitudes at hand.
To calculate the temperature, we used the  
color calibrations from  \citet{Alonso1999}, which are based on the infrared flux method (IRFM)  for giants.
Furthermore,  ($J-K$) is  the color of the three IR colors with the lowest internal uncertainty \citep{Alonso1996,Alonso1999}. 
{Moreover, these authors also recommended the use of ($J-K$) when the metallicities were uncertain.}
We adopted the mean E(B$-$V) from \citet{Schlafly2011} and  converted extinction coefficients A$_J$ and A$_K$ from the IRSA Dust
database\footnote{http://irsa.ipac.caltech.edu/applications/DUST/}
to deredden the $J-K$-colors. 
{
To calculate T$_{\rm eff}$, we used the SP\_ACE output for [Fe/H].  
Owing to the null metallicity dependence in the IRFM ($J-K$) color calibrations of \citet{Alonso1996,Alonso1999}, we obtain identical temperatures (within 1--2 K) 
if we were to assume a fixed value of $-$1 dex versus the metallicities computed with SP\_ACE, unless
the adopted metallicity would push the star outside the allowed metallicity bin in the calibration. 
Therefore, we did not further iterate the calculated temperatures because of uncertainties in [Fe/H]. 
We note that seven stars deviate strongly from the otherwise good correlation between the Alonso and SP\_ACE temperatures. 
These  stars fall outside the  metallicity range covered by the IRFM T$_{\rm eff}$-color calibration. Above 4800 K, approximately 30 stars yield  
Alonso temperatures considerably lower than those derived by SP\_ACE. 
These stars are warmer, but faint, and the vast majority of them have reddening values above 0.4 mag, 
and they are most likely poorly classified; we also note a slight increase in the temperature difference with increasing gravity.
}
%For calculating T$_{\rm eff}$, we assumed a fixed metallicity of [Fe/H]$=-1$ (in agreement with the values derived by SP\_ACE). 
%This is a good assumption for the cluster members and a solid mid-range value for the foreground stars according to Fig.~4. 
%Owing to the null metallicity dependence in the IRFM $J-K$ colour calibration of \citet{Alonso1999},  we obtain identical 
%temperatures for all metallicities considered. Therefore, we did not iterate  the calculated temperatures due to uncertainties in [Fe/H].

Following this, we tested the empirical log\,$g$-T$_{\rm eff}$-relation from \citet{Barklem2005}  to estimate the gravities. 
However, their empirical formula  was calibrated based on stars
that are typically warmer than 4300K, while many of our targets are very cool objects ($<4300$K), 
 resulting in highly negative gravities. Thus we discarded this approach and continued using the SP\_ACE parameters described above.
\subsection{Final stellar parameters}
Figure~4 shows a Hertzsprung-Russel diagram of the entire target sample, color-coded by their iron abundance from SP\_ACE. 
The RGB is well populated down to its base at log\,$g$$\sim$3.5 and the T$_{\rm eff}$-metallicity relation is indicated by 
a dozen metal-poor warmer giants. Furthermore, the foreground component contains a few metal-rich dwarf stars, typical of the Galactic disks. 
As a result of our color-selection criteria from the 2MASS, the majority of stars in the sample are therefore K-giants 
with temperatures lower than $\sim$5000 K.
\begin{figure}[htb]
\centering
\includegraphics[width=1\hsize,  clip=true]{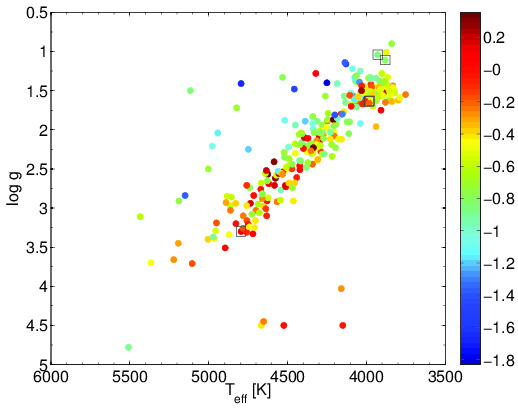}
\caption{Hertzsprung-Russell diagram of ESO452-SC11 with stellar parameters from SP\_ACE, color-coded by iron abundance.
The member candidates are highlighted with black squares. Note that two candidates with very similar stellar parameters overlap in this diagram.}
\end{figure}

We note that one of our GC member candidates (number 56) has a considerably higher log\,$g$, at $\sim$3.3 dex, compared to the 
values found for  the other four,  which lie at $\sim$1--1.6 dex. 
While there is no reason why our selection criteria should not have picked stars at the base of the RGB, its
deviating chemical abundance ratios indicate that this object may be a foreground star as well.
\begin{figure}[htb]
\centering
\includegraphics[width=1\hsize,  clip=true]{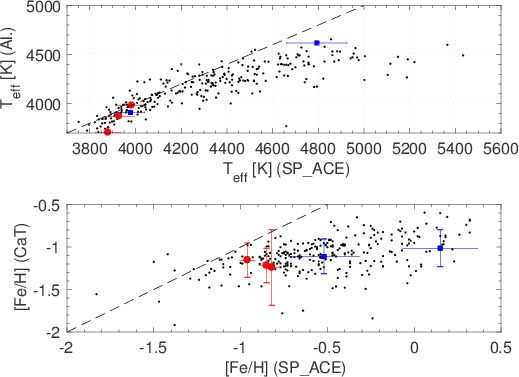}
\caption{Comparison of stellar parameters that were obtained through different methods. Red symbols show the innermost three member candidates, while 
blue symbols denote the two outer candidates with less secure membership. The dashed line shows unity.}
\end{figure}
In Fig.~5 we compare the stellar temperatures obtained with the different methods outlined above.
The calibrations of Alonso return T$_{\rm eff}$ that are {62 K lower on average than the SP\_ACE results, with a scatter of 109 K below
4500 K. Above this temperature, the agreement worsens for the reasons discussed above.} 
We note that the five GC candidates are in very good agreement with the Alonso scale.
For the remainder of our analysis we  adopt the stellar parameters from SP\_ACE and note that although it is constrained to the short-wavelength coverage of the red AAOMega fibres, the code returns reliable values for T$_{\rm eff}$ and more physically meaningful values for gravities and metallicities than any scaling formula we tested. 
\subsection{Iron abundance}
In Fig.~5 (bottom panel) we compare the metallicities from the CaT with those obtained by the direct spectral analysis within SP\_ACE. 
As noted above, CaT metallicities are meaningless for the foreground stars because the calibration cannot be applied for want of 
a reference red clump magnitude.
The mean difference of the metallicity for the five GC member candidates from the Fe-lines and their CaT counterparts is 0.55 dex 
with a 1$\sigma$ scatter of 0.37 dex.
Here, the CaT metallicities give systematically lower values and the difference is significant at the 1--3$\sigma$ level. 
In order to reconcile the measurements for the member stars, the RC magnitude would need to be lowered by $\sim$1.5 magnitudes; reddening uncertainties 
on that order of magnitude are unlikely. 
These differences, however, are driven by the two objects that are located farthest away from the GC center (stars 56 and 396), and as we show  in Sect.~4.5. below, 
the higher-gravity object 56 has a high Fe-abundance of 0.15$\pm$0.08 dex and 
also solar-scaled [$\alpha$/Fe] abundance ratios, suggesting that these two deviating objects are most likely  foreground contaminants as well. 

The mean Fe-abundance of the three {\em \textup{bona fide}} GC member candidates (star IDs 55, 204, and 362)  
is $-0.88\pm 0.03$ dex with a metallicity dispersion of $0.03\pm0.04$ dex; this is higher than the value predicted by the CaT by 0.31 dex.
As for the radial velocity dispersion of this system, also the very low spread in Fe-abundances is fully compatible with the low mass of 
ESO452 \citep{Carretta2009,Koch2012}.

Figure~6  in turn shows the MDF from our entire sample, ranging from the metal-poor regime ([Fe/H]$\sim -1.8$) to slightly supersolar values of $\sim$0.3 dex.
In that regard, the field surrounding ESO452 shows the typical shape and range for its location toward the Galactic bulge, including contributions from 
all nearby major subcomponents, including an inner halo, metal-weak thick disk, bulge and disk stars \citep{Kunder2012,Ness2013,Zoccali2017}.
The MDF, however, lacks the pronounced peak at solar values with its strong tail far above solar metallicities, as is seen in low-latitude fields
that are uniquely bulge-dominated. 
\begin{figure}[htb]
\centering
\includegraphics[width=0.8\hsize, clip=true]{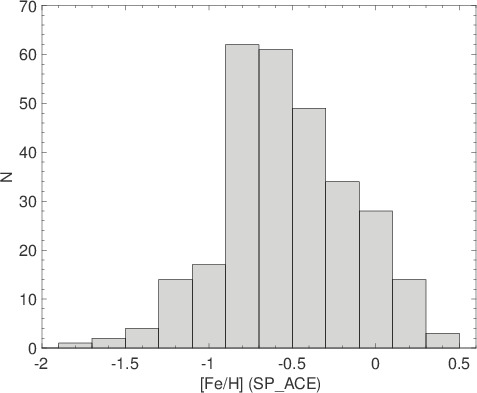}
\caption{Metallicity distribution of the entire foreground plus GC sample, using the Fe-abundances derived from SP\_ACE.}
\end{figure}
\subsection{Alpha-elements:  Si, Ca, and Ti}
In Fig.~7 we compare the measurements of our GC and foreground stars with samples from the literature. 
Since the errors on our data are considerably large, we chose to consider in the following 
the  straight average of the three explosive $\alpha$-elements Si, Ca, and Ti as a representative chemical tracer, even though this simplistic approach cannot account for 
differences in their exact production channels \citep[e.g.,][]{Timmes1995,Venn2004}. 
\begin{figure}[htb]
\centering
\includegraphics[width=1\hsize, clip=true]{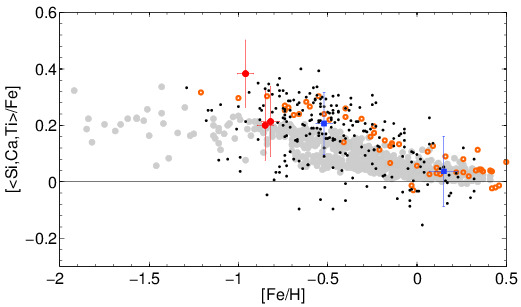}
\caption{Averaged $\alpha$-element abundances for our sample (black dots) in comparison with stars in the solar neighborhood  from \citet{Bensby2014} (light gray shade) and in 
the Galactic bulge \citep{Bensby2013}, shown in orange.  
The best cluster candidates, selected on grounds of their kinematics and position, are highlighted in red, while
blue symbols denote the two stars with GC properties except for deviating abundances. }
\end{figure}

For comparison, we use the  solar neighborhood sample of \citet[][light gray points]{Bensby2014} and microlensed bulge dwarfs of \citet[][orange]{Bensby2013}.
There is a clear dichotomy in the data in the sense of the bulge stars having 
 [$\alpha$/Fe] ratios that are higher by 0.1--0.2 dex at subsolar metallicities, thereby vastly overlapping with the thick disk \citep[see also][]{Fulbright2007,McWilliam2016}. 
Taken at face value, the majority of our non-cluster stars at their higher $\alpha$-abundances 
are therefore more consistent with an association with the bulge population than with the disk.
\citet{Fulbright2007} assigned these values to indications of supernovae II nucleosynthesis and pointed out the considerably lower scatter in the averaged [$<$Si,Ca,Ti$>$/Fe]
ratio compared to the Galactic halo, for example, which suggests efficient mixing in the bulge.

 About a third of the sample shows lower [$\alpha$/Fe] ratios
that overlap with the solar neighborhood stars, down to values below zero. This renders the stars toward the sightline of ESO452 a 
representative mix of the expected components \citep[see also][]{Koch2016}. 

Two of the  most ``metal-poor'' stars among the five GC candidates have slightly depleted  [$\alpha$/Fe] ratios, while the third star is enhanced to halo-like values, 
resulting in an average for the three stars of 0.27 dex with a  1$\sigma$ scatter of
0.10 dex.
This is compatible with the disk population and does not show the elevated levels of the bulge, although the large error bars 
make this distinction marginal. 
One star at [Fe/H]=$-0.5$ dex shares this abundance, but its higher metallicity renders it a likely foreground object.
Finally, the solar-scaled [$\alpha$/Fe] ratio of most metal-rich star (number 56) confirms that it is a typical metal-rich disk star without any association with ESO452. 
\subsection{Fe-peak elements: Cr, Co, and Ni}
The measurement of these abundances from just a few lines at most  is very uncertain, and the resulting distribution (Fig.~8)
of our stars show mainly a large scatter around the solar values, for both the foreground stars and the GC candidates.
In contrast, the literature data from the high-resolution studies in the disk and bulge show a very low scatter and broadly overlap with each other.
Owing to the large uncertainties and scatter of our measurements, we refrain from any strong conclusions on the enrichment of ESO452 in 
comparison to the Galactic components.
\begin{figure}[htb]
\centering
\includegraphics[width=1\hsize, clip=true]{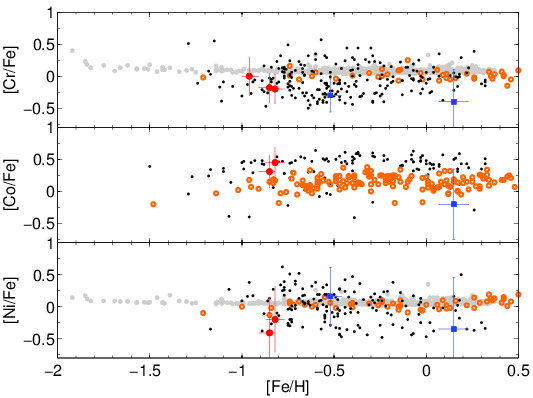}
\caption{Same as Fig.~7, but for the Fe-peak elements. Here, the Co-data are from \citet{Johnson2014}.}
\end{figure}
\subsection{Neutron-capture elements: Sr and Eu}
{The blue part of the AAOmega spectra is plagued by low S/N and most features are heavily blended. Thus we caution
 that the derived abundances are uncertain, and we impose a fitting uncertainty of $\pm0.15$ dex that is due to continuum placement and blends alone.
 Moreover, since the heavy blending at the moderate resolution rendered measurements of abundances of the metal-rich foreground component impossible, we restricted our analysis
 of the blue spectra to the five GC member candidates. 

The Sr abundance was derived from the blue spectra by virtue of the strong 4077 and 4215 \AA~lines using spectral synthesis; since the spectral range also covered the wavelength 
of the two strongest blue Eu lines, we also attempted to derive abundances or place limits from the features at 4129 and 4205 \AA. 

To this end, we used Moog \citep[][version 2014]{Sneden1973} and Kurucz model atmospheres adopting the SP\_ACE stellar parameters derived above. 
The line lists were optimized for the blue wavelength range using the \citet{Sneden2014} list that incorporates the newest hyperfine structure data for Sr 
\citep{Bergemann2012,CJHansen2013}.
In order to treat the continuum best, the blue spectra were normalized in two different ways 
using a cubic spline and Legendre polynomial. 
For star 204, which has the best S/N, this was straightforward. However, for stars 396 and 55, we adopted the average abundance from each of the two normalizations, which yielded 
results that agreed to within 0.1 dex. 
To further optimize the continuum placement, the synthesis was carried out using the full spectral range in  pieces 100 \AA~wide each 
(4050--4150 \AA~and 4150--4250 \AA~around the two Sr and Eu lines). 
This allowed us to derive Sr abundances in three stars, Eu in star 204, and one  lower limit for Eu in star 396 due to the low S/N and line saturation.  
A section of of our observations and the best fit is shown in Fig.~9.
\begin{figure}[htb]
\centering
%\hspace{0.4cm}\includegraphics[width=0.85\hsize, clip=true]{f9a.jpg}
%\includegraphics[width=1\hsize, clip=true]{f9b.jpg}
\includegraphics[width=1\hsize, clip=true]{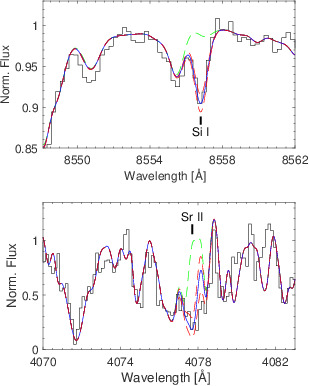}
\caption{Synthetic spectra around the Si~8556 \AA-line (top) and the Sr~4077 \AA-line (bottom panel) in star 204 (black {histograms}). The blue lines 
have been computed with the best-fit abundances of [Si/Fe]=0.44  and [Sr/Fe]=0.00 dex, accounting for errors of $\pm$0.2 dex (red). 
The green lines show syntheses without any Si and Sr, respectively.}
\end{figure}

As can bee seen in Fig.~10, the derived Sr abundances are  above solar and are found to be in agreement with the disk stars from \citet{Battistini2016} and the 
mixed halo and disk sample of \citet{CJHansen2012}. Data for Sr in the Galactic bulge are only available for a few metal-poor
field stars below $-$2 dex \citep{CaseySchlaufman2015,Howes2016,Koch2016}, and our measurements provide the first Sr abundances in 
a metal-rich system toward the bulge.
\begin{figure}[htb]
\centering
\includegraphics[width=1\hsize, clip=true]{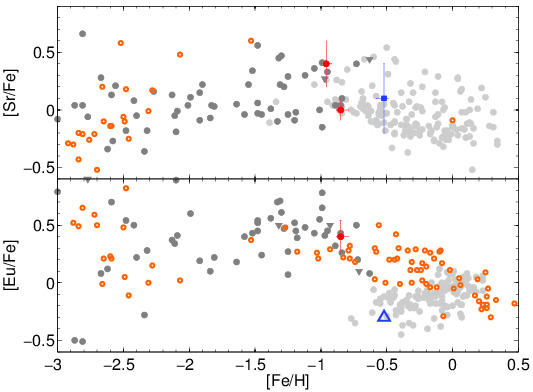}
\caption{Abundance ratios for the $n$-capture elements Sr and Eu with literature data.
As before, gray points denote disk and halo stars of \citet[][dark gray]{CJHansen2012}  and  \citet[][light gray]{Battistini2016} for Sr, and 
\citet[][dark gray]{CJHansen2012} and \citet[][light gray]{Koch2002} for Eu. Orange dots show the bulge component \citep{Johnson2012}, extending to the metal-poor bulge below [Fe/H] $< -2$ dex \citep{CaseySchlaufman2015,Howes2016,Koch2016}.}
\end{figure}

The Eu abundance for one GC star is in turn compatible with the bulge population \citep{Johnson2012}, while the less reliable candidate, at [Fe/H]=$-0.5$ dex,
has a lower limit in agreement with either bulge or disk \citep{Koch2002}. In  light of the measurement uncertainties, we refrain from unambiguously assigning these targets to either Galactic 
component. 
\section{Correlation with kinematics}
For an understanding of the origin and dynamical evolution of ESO452, it is of considerable interest to determine
the orbit of this GC.  However, at the distance of ESO452, it is nontrivial to
obtain proper motions of sufficient quality for a direct calculation of the orbit.  Nevertheless, 
we cross-matched the stars observed here with the
HSOY (``Hot Stuff for One Year'') proper motion catalog \citep{Altmann2017} as well as the UCAC5 
proper motion catalog \citep{Zacharias2017}.  Both these catalogs capitalize on the exquisite positions
provided in Gaia-DR1 \citep{GaiaDR1},
providing  proper motions much improved over the previously available catalogs.  

We were able to match 341 of our observed stars (94\% of our sample) with those provided in
HSOY and UCAC5,                 % Out of the 400 fibres, 39 are sky / parked / guide, so only 361 are 'real' stars. 341 / 361 = 94%
where two of these are probable GC members.
Figure~11 shows that the proper motions of the two GC member candidates that are present in the catalog 
are slightly offset from
the majority of the bulge field stars, in agreement with the
finding that our GC candidates are distinct from the field and
instead belong to a different system.  The two candidate cluster stars have HSOY proper motions that 
differ by 2--3$\sigma$ from each other, indicating that if these two stars are indeed cluster members,
the uncertainties in their proper motion should be larger than indicated in HSOY.  In contrast, the UCAC5
proper motions, which have smaller formal uncertainties, indicate that our cluster candidates have almost 
identical proper motions with an average of ($\rm \mu_{\alpha} , \mu_{\delta}$) = $(-0.6 \pm 1.1, -6.0 \pm 1.2)$ mas.  We therefore conclude that the proper motions also indicate that our cluster candidates are members of ESO452.
\begin{figure}[htb]
\centering
\includegraphics[width=1\hsize]{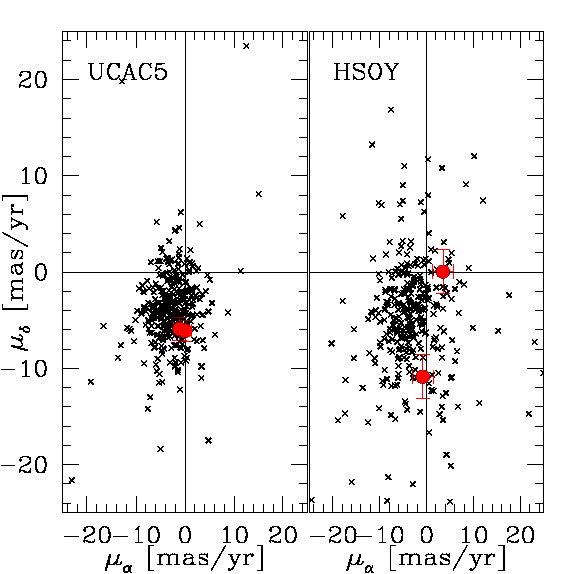}
\includegraphics[width=1\hsize]{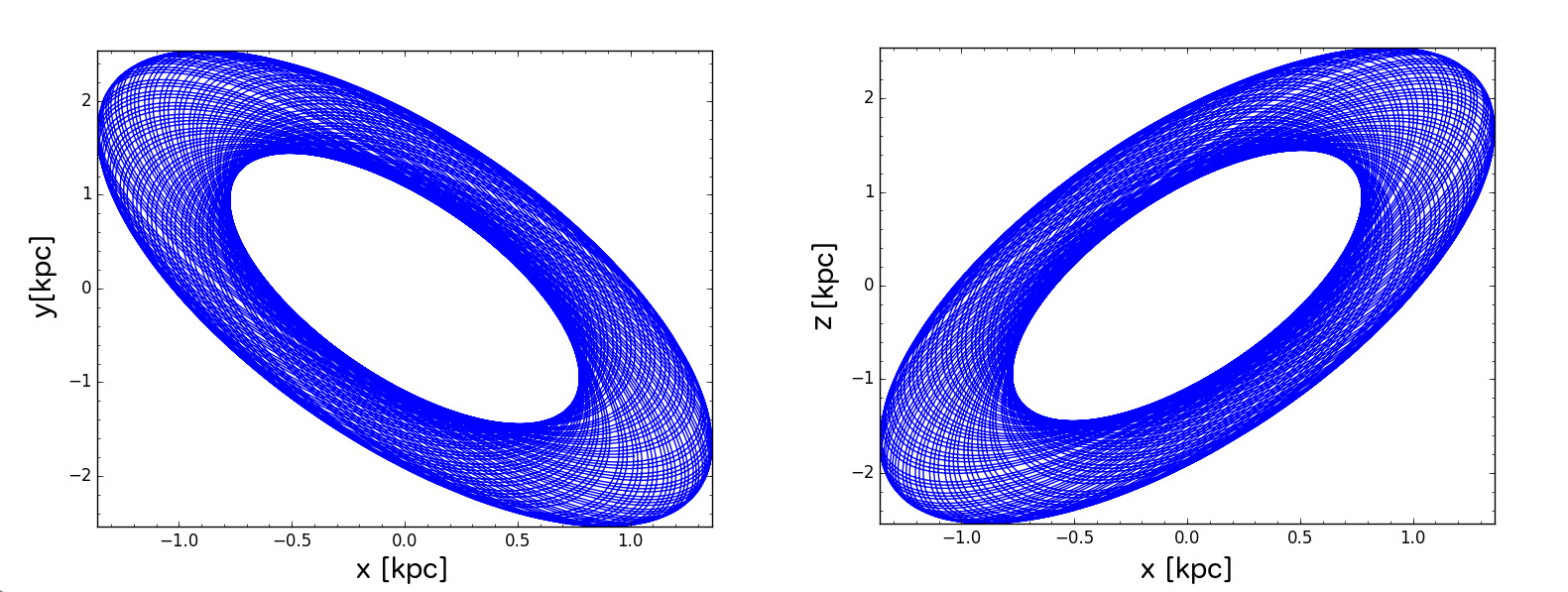}
\caption{{\sl Top:} The proper motions of our observed stars, where the solid red circles show the two GC candidates that 
are listed in both the HSOY and UCAC5 catalogs. 
{\sl Bottom:} Projections of the orbit of the two candidate ESO452 stars in the xy and xz plane.  The orbits
indicate that they are confined to the bulge.
}
\end{figure}

Adopting our radial velocities, the Gaia positions, the UCAC5 proper
motions, and a distance of 7.2~kpc \citep{Cornish2006}, 
we calculated orbits using the python package {\tt galpy} \citep{Bovy2015}.  
The Galactic potential assumed was a three-component Milky Way-like potential for the orbital integration.  
The two candidate cluster members have orbits with ellipticities of $e$$\sim$0.3, a maximum vertical height, $Z_{max}$, 
of $\sim$2.5~kpc, and orbital pericenter ($r_{min}$) and apocenter ($r_{max}$) radii of 
$\sim$1.5~kpc and $\sim$3~kpc, respectively, placing them well within the bulge region.
Adopting a 1~kpc larger cluster distance does not change the results significantly.
In addition, using the slightly different proper motions of HSOY also returns orbits in which the
stars are confined to the bulge.  The orbital integration therefore
indicates that this GC belongs to the bulge, 
in line with its abundance and velocity.

Turning to the Galactocentric velocity, this value corresponds to 
  $-4.3\pm1.1$ km\,s$^{-1}$ \citep[as determined by Eq. 1 in][]{Howard2008}.
While this is a higher velocity than the mean stellar motion of the outer bulge \citep{Kunder2012},  it
 falls well within the range of velocities of GCs that belong to the bulge \citep[][and Fig.~12, top]{Bica2016}. In terms of its metallicity, 
 it also shows a mean value that agrees with the values found among the bulge GCs, which is expected because they cover a range of more than 1 dex  
 in [Fe/H].
The mean Galactocentric velocity of our entire sample is $-60.7\pm4.7$ km\,s$^{-1}$ with a 1$\sigma$ dispersion of 
79.7$\pm$3.3  km\,s$^{-1}$.  
These values are well in line with what is expected of bulge field stars
when we extrapolate the southern bulge rotation curve (e.g., Kunder et~al. 2012, Ness et~al. 2013) 
to this northern position on the sky (Fig.~12). 
\begin{figure}[htb]
\centering
\includegraphics[width=1\hsize, clip=true]{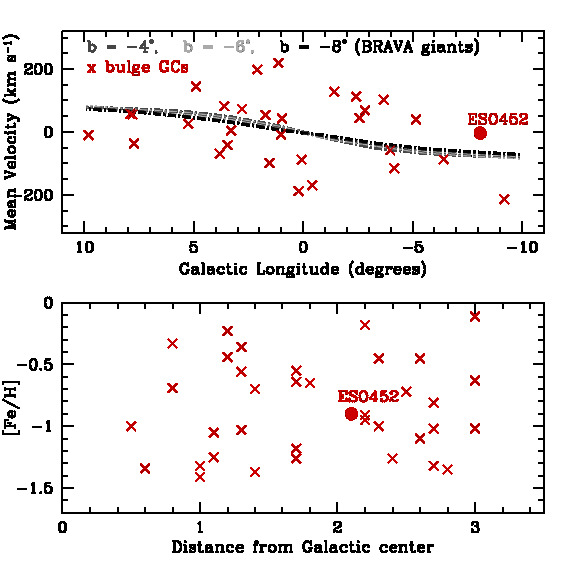}
\caption{{\sl Top:} The rotation curve for bulge stars \citep{Kunder2012} and that of  bulge GCs \citep[][]{Bica2016}.
{\sl Bottom:} Metallicities of  bulge globular clusters.  
Overall, ESO452 has properties that are   
expected for a typical system in the Galactic bulge.}
\end{figure}
\section{No extra-tidal stars}
Our chemodynamical sample offers a unique opportunity of testing for signs of extra-tidal stars, which could have been stripped from the GC
through tidal interactions while orbiting within the Galactic potential \citep[e.g,][]{Kunder2014Tidal,Kuzma2016}. 
To this end, we imposed a 3$\sigma$ cut around the cluster mean radial velocity 
to select stars outside the nominal tidal radius that share similar properties as the three {\em \textup{bona fide}} member candidates characterized above
(stars 55, 204, and 362).

\begin{figure}[htb]
\centering
\includegraphics[width=0.8\hsize, clip=true]{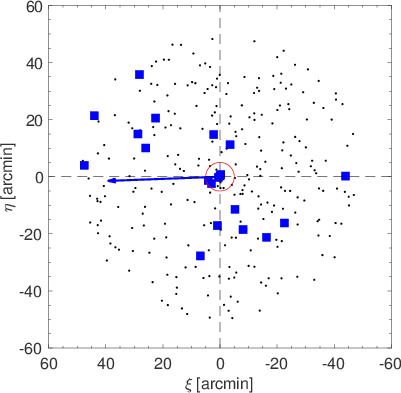}
\caption{Spatial location of all sample stars (black) and targets within 3$\sigma$ of the mean velocity of the GC candidates (blue squares).
The blue arrow indicates the direction toward the Galactic center, and the red circle encompasses the nominal tidal radius.}
\end{figure}
As a result, we identify 15 stars within this kinematic cut in addition to the candidates within the tidal radius discussed above. 
Their location within our observations' field of view is shown in Fig.~13. 
As it stands, this would imply that these objects are preferentially aligned symmetrically around the GC, NW through SE. 
If these stars were indeed to represent a tidal extension of the GC, they should roughly follow the GC orbit around the Galactic center
\citep[e.g.,][]{Dehnen2004,Jordi2010}. However, as indicated in Fig.~13, no obvious correlation with direction of the Milky
Way center can 
be established. 
Furthermore, these 15 stars show no coherence in proper motions or abundance space and are instead found across the entire range in proper motion, metallicity, and [$\alpha$/Fe]-ratios with no overlap
with the three best GC candidates. We conclude that our measured velocities and abundances offer no evidence of tidally stripped material from ESO452. 
\section{Discussion}
In contrast to remote and uniquely metal-poor GCs in the Galactic halo, stellar systems in the interface of the Milky Way disk,
bulge, and halo are 
notoriously difficult to classify. By focusing on the chemical composition and dynamics of one of these elusive systems, the low-mass GC ESO452-SC11, 
we were able to gather significant information to place it in
the context of the central regions of our Galaxy. 
Considering its moderately low metallicity of $-0.88$ dex and a {Galactocentric velocity of $-4$ km\,s$^{-1}$,} and  folding in the 
chemical abundance ratios we obtained, we could establish strong evidence that this object is indeed associated with 
the Galactic bulge and is not a somewhat younger progeny of the inner halo. 
By identifying at least three {\em \textup{bona fide}} GC member candidates, we showed that their [$\alpha$/Fe] ratios are  consistent with a disk membership as well,
while they are chemically marginally consistent with the slightly $\alpha$-enhanced bulge component. 
To be of avail, future high-resolution studies of proper motion pre-selected targets are indispensable. 

{
ESO452 is a very good example of a low-mass metal-intermediate ([Fe/H]$\sim -1$ dex) GC that resembles for instance the inner halo 
clusters NGC 6362 and NGC 6121 \citep{Marino2008,Massari2017} in many regards. These are among the lowest-mass systems that 
show the presence of multiple stellar populations through light element variations, which remain undetected in ESO452 in our work, however.
Still, the $\alpha$-element, Cr, and Eu abundances in the bulge GC studied here agree well with those found in the above objects, and their
other elemental abundances are also consistent within the uncertainties.}
}

By considering the proper motions of the candidate members, we were able to consolidate a clear connection with the Galactic bulge in that 
this GC exhibits an elliptical orbit that confines it to the innermost $\sim$3 kpc of the Galaxy. 
Despite the higher stellar density of the inner Galactic region, no evidence of any tidal happenings  in the form 
of stars beyond the tidal radius were found in our wide-field data, which either indicates  that any spatial signatures could have been erased
during violent GC disruption, or that our data simply have not picked up any such ``extra-tidal'' component. The former is bolstered by the 
 lack of kinematically cold streams in the bulge \citep{Kunder2014Streams}, however, which would be witnesses to ancient accretion events \citep[cf.][]{Hansen2016RR}, 
 as has also been proposed 
 for ESO452 \citep{Cornish2006} and as has indeed been found around several other systems \citep{Kunder2014Tidal}. 
\begin{acknowledgements}
Based on data acquired through the Australian Astronomical Observatory, via program NOAO Prop. ID: 2013A-0120 (PI: A. Kunder).
C.J. Hansen acknowledges financial support by the Augustinus Foundation and 
 the VILLUM Foundation (grant number VKR023371).
We are grateful to the anonymous referee for a swift and constructive report.  
 \end{acknowledgements}
\small
\bibliographystyle{aa} % style aa.bst
\bibliography{30788_ap_AK} % your references Yourfile.bib

\begin{thebibliography}{70}
\expandafter\ifx\csname natexlab\endcsname\relax\def\natexlab#1{#1}\fi

\bibitem[{{Alonso} {et~al.}(1996){Alonso}, {Arribas}, \&
  {Martinez-Roger}}]{Alonso1996}
{Alonso}, A., {Arribas}, S., \& {Martinez-Roger}, C. 1996, \aap, 313, 873

\bibitem[{{Alonso} {et~al.}(1999){Alonso}, {Arribas}, \&
  {Mart{\'{\i}}nez-Roger}}]{Alonso1999}
{Alonso}, A., {Arribas}, S., \& {Mart{\'{\i}}nez-Roger}, C. 1999, \aaps, 140,
  261

\bibitem[{{Altmann} {et~al.}(2017){Altmann}, {Roeser}, {Demleitner}, {Bastian},
  \& {Schilbach}}]{Altmann2017}
{Altmann}, M., {Roeser}, S., {Demleitner}, M., {Bastian}, U., \& {Schilbach},
  E. 2017, \aap, 600, L4

\bibitem[{{Armandroff} \& {Zinn}(1988)}]{Armandroff1988}
{Armandroff}, T.~E. \& {Zinn}, R. 1988, \aj, 96, 92

\bibitem[{{Barklem} {et~al.}(2005){Barklem}, {Christlieb}, {Beers}, {Hill},
  {Bessell}, {Holmberg}, {Marsteller}, {Rossi}, {Zickgraf}, \&
  {Reimers}}]{Barklem2005}
{Barklem}, P.~S., {Christlieb}, N., {Beers}, T.~C., {et~al.} 2005, \aap, 439,
  129

\bibitem[{{Battistini} \& {Bensby}(2016)}]{Battistini2016}
{Battistini}, C. \& {Bensby}, T. 2016, \aap, 586, A49

\bibitem[{{Bensby} {et~al.}(2014){Bensby}, {Feltzing}, \& {Oey}}]{Bensby2014}
{Bensby}, T., {Feltzing}, S., \& {Oey}, M.~S. 2014, \aap, 562, A71

\bibitem[{{Bensby} {et~al.}(2013){Bensby}, {Yee}, {Feltzing}, {Johnson},
  {Gould}, {Cohen}, {Asplund}, {Mel{\'e}ndez}, {Lucatello}, {Han}, {Thompson},
  {Gal-Yam}, {Udalski}, {Bennett}, {Bond}, {Kohei}, {Sumi}, {Suzuki}, {Suzuki},
  {Takino}, {Tristram}, {Yamai}, \& {Yonehara}}]{Bensby2013}
{Bensby}, T., {Yee}, J.~C., {Feltzing}, S., {et~al.} 2013, \aap, 549, A147

\bibitem[{{Bergemann} {et~al.}(2012){Bergemann}, {Hansen}, {Bautista}, \&
  {Ruchti}}]{Bergemann2012}
{Bergemann}, M., {Hansen}, C.~J., {Bautista}, M., \& {Ruchti}, G. 2012, \aap,
  546, A90

\bibitem[{{Bica} {et~al.}(2016){Bica}, {Ortolani}, \& {Barbuy}}]{Bica2016}
{Bica}, E., {Ortolani}, S., \& {Barbuy}, B. 2016, \pasa, 33, e028

\bibitem[{{Boeche} \& {Grebel}(2016)}]{Boeche2016}
{Boeche}, C. \& {Grebel}, E.~K. 2016, \aap, 587, A2

\bibitem[{{Bovy}(2015)}]{Bovy2015}
{Bovy}, J. 2015, \apjs, 216, 29

\bibitem[{{Carretta} {et~al.}(2009){Carretta}, {Bragaglia}, {Gratton},
  {D'Orazi}, \& {Lucatello}}]{Carretta2009}
{Carretta}, E., {Bragaglia}, A., {Gratton}, R., {D'Orazi}, V., \& {Lucatello},
  S. 2009, \aap, 508, 695

\bibitem[{{Casey} \& {Schlaufman}(2015)}]{CaseySchlaufman2015}
{Casey}, A.~R. \& {Schlaufman}, K.~C. 2015, \apj, 809, 110

\bibitem[{{Chiba} \& {Beers}(2000)}]{ChibaBeers2000}
{Chiba}, M. \& {Beers}, T.~C. 2000, \aj, 119, 2843

\bibitem[{{Cornish} {et~al.}(2006){Cornish}, {Phelps}, {Briley}, \&
  {Friel}}]{Cornish2006}
{Cornish}, A.~S.~M., {Phelps}, R.~L., {Briley}, M.~M., \& {Friel}, E.~D. 2006,
  \aj, 131, 2543

\bibitem[{{Cutri} {et~al.}(2003){Cutri}, {Skrutskie}, {van Dyk}, {Beichman},
  {Carpenter}, {Chester}, {Cambresy}, {Evans}, {Fowler}, {Gizis}, {Howard},
  {Huchra}, {Jarrett}, {Kopan}, {Kirkpatrick}, {Light}, {Marsh}, {McCallon},
  {Schneider}, {Stiening}, {Sykes}, {Weinberg}, {Wheaton}, {Wheelock}, \&
  {Zacarias}}]{Cutri2003}
{Cutri}, R.~M., {Skrutskie}, M.~F., {van Dyk}, S., {et~al.} 2003, {2MASS All
  Sky Catalog of point sources.}

\bibitem[{{Dehnen} {et~al.}(2004){Dehnen}, {Odenkirchen}, {Grebel}, \&
  {Rix}}]{Dehnen2004}
{Dehnen}, W., {Odenkirchen}, M., {Grebel}, E.~K., \& {Rix}, H.-W. 2004, \aj,
  127, 2753

\bibitem[{{Dotter} {et~al.}(2008){Dotter}, {Chaboyer}, {Jevremovi{\'c}},
  {Kostov}, {Baron}, \& {Ferguson}}]{Dotter2008}
{Dotter}, A., {Chaboyer}, B., {Jevremovi{\'c}}, D., {et~al.} 2008, \apjs, 178,
  89

\bibitem[{{Dubath} {et~al.}(1997){Dubath}, {Meylan}, \& {Mayor}}]{Dubath1997}
{Dubath}, P., {Meylan}, G., \& {Mayor}, M. 1997, \aap, 324, 505

\bibitem[{{Ferraro} {et~al.}(2009){Ferraro}, {Dalessandro}, {Mucciarelli},
  {Beccari}, {Rich}, {Origlia}, {Lanzoni}, {Rood}, {Valenti}, {Bellazzini},
  {Ransom}, \& {Cocozza}}]{Ferraro2009}
{Ferraro}, F.~R., {Dalessandro}, E., {Mucciarelli}, A., {et~al.} 2009, \nat,
  462, 483

\bibitem[{{Fulbright} {et~al.}(2007){Fulbright}, {McWilliam}, \&
  {Rich}}]{Fulbright2007}
{Fulbright}, J.~P., {McWilliam}, A., \& {Rich}, R.~M. 2007, \apj, 661, 1152

\bibitem[{{Hansen} {et~al.}(2012){Hansen}, {Primas}, {Hartman}, {Kratz},
  {Wanajo}, {Leibundgut}, {Farouqi}, {Hallmann}, {Christlieb}, \&
  {Nilsson}}]{CJHansen2012}
{Hansen}, C.~J., {Primas}, F., {Hartman}, H., {et~al.} 2012, \aap, 545, A31


\bibitem[{{Hansen} {et~al.}(2013){Hansen}, {Bergemann}, {Cescutti}, {Fran{\c
  c}ois}, {Arcones}, {Karakas}, {Lind}, \& {Chiappini}}]{CJHansen2013}
{Hansen}, C.~J., {Bergemann}, M., {Cescutti}, G., {et~al.} 2013, \aap, 551, A57

\bibitem[{{Hansen} {et~al.}(2015){Hansen}, {Ludwig}, {Seifert}, {Koch}, {Xu},
  {Caffau}, {Christlieb}, {Korn}, {Lind}, {Sbordone}, {Ruchti}, {Feltzing}, {de
  Jong}, \& {Barden}}]{CJHansen2015}
{Hansen}, C., {Ludwig}, H.-G., {Seifert}, W., {et~al.} 2015, Astronomische
  Nachrichten, 336, 665

\bibitem[{{Hansen} {et~al.}(2016){Hansen}, {Rich}, {Koch}, {Xu}, {Kunder}, \&
  {Ludwig}}]{Hansen2016RR}
{Hansen}, C.~J., {Rich}, R.~M., {Koch}, A., {et~al.} 2016, \aap, 590, A39

\bibitem[{{Harris}(1996)}]{Harris1996}
{Harris}, W.~E. 1996, \aj, 112, 1487

\bibitem[{{Hendricks} {et~al.}(2014){Hendricks}, {Koch}, {Lanfranchi},
  {Boeche}, {Walker}, {Johnson}, {Pe{\~n}arrubia}, \&
  {Gilmore}}]{Hendricks2014}
{Hendricks}, B., {Koch}, A., {Lanfranchi}, G.~A., {et~al.} 2014, \apj, 785, 102

\bibitem[{{Howard} {et~al.}(2008){Howard}, {Rich}, {Reitzel}, {Koch}, {De
  Propris}, \& {Zhao}}]{Howard2008}
{Howard}, C.~D., {Rich}, R.~M., {Reitzel}, D.~B., {et~al.} 2008, \apj, 688,
  1060

\bibitem[{{Howes} {et~al.}(2016){Howes}, {Asplund}, {Keller}, {Casey}, {Yong},
  {Lind}, {Frebel}, {Hays}, {Alves-Brito}, {Bessell}, {Casagrande}, {Marino},
  {Nataf}, {Owen}, {Da Costa}, {Schmidt}, \& {Tisserand}}]{Howes2016}
{Howes}, L.~M., {Asplund}, M., {Keller}, S.~C., {et~al.} 2016, \mnras, 460, 884

\bibitem[{{Jofr{\'e}} {et~al.}(2014){Jofr{\'e}}, {Heiter}, {Soubiran},
  {Blanco-Cuaresma}, {Worley}, {Pancino}, {Cantat-Gaudin}, {Magrini},
  {Bergemann}, {Gonz{\'a}lez Hern{\'a}ndez}, {Hill}, {Lardo}, {de Laverny},
  {Lind}, {Masseron}, {Montes}, {Mucciarelli}, {Nordlander}, {Recio Blanco},
  {Sobeck}, {Sordo}, {Sousa}, {Tabernero}, {Vallenari}, \& {Van
  Eck}}]{Jofre2014}
{Jofr{\'e}}, P., {Heiter}, U., {Soubiran}, C., {et~al.} 2014, \aap, 564, A133

\bibitem[{{Johnson} {et~al.}(2012){Johnson}, {Rich}, {Kobayashi}, \&
  {Fulbright}}]{Johnson2012}
{Johnson}, C.~I., {Rich}, R.~M., {Kobayashi}, C., \& {Fulbright}, J.~P. 2012,
  \apj, 749, 175

\bibitem[{{Johnson} {et~al.}(2014){Johnson}, {Rich}, {Kobayashi}, {Kunder}, \&
  {Koch}}]{Johnson2014}
{Johnson}, C.~I., {Rich}, R.~M., {Kobayashi}, C., {Kunder}, A., \& {Koch}, A.
  2014, \aj, 148, 67

\bibitem[{{Jordi} \& {Grebel}(2010)}]{Jordi2010}
{Jordi}, K. \& {Grebel}, E.~K. 2010, \aap, 522, A71

\bibitem[{{Kleyna} {et~al.}(2004){Kleyna}, {Wilkinson}, {Evans}, \&
  {Gilmore}}]{Kleyna2004}
{Kleyna}, J.~T., {Wilkinson}, M.~I., {Evans}, N.~W., \& {Gilmore}, G. 2004,
  \mnras, 354, L66

\bibitem[{{Koch} \& {Edvardsson}(2002)}]{Koch2002}
{Koch}, A. \& {Edvardsson}, B. 2002, \aap, 381, 500

\bibitem[{{Koch} \& {McWilliam}(2008)}]{Koch2008}
{Koch}, A. \& {McWilliam}, A. 2008, \aj, 135, 1551

\bibitem[{{Koch} {et~al.}(2009){Koch}, {C{\^o}t{\'e}}, \&
  {McWilliam}}]{Koch2009}
{Koch}, A., {C{\^o}t{\'e}}, P., \& {McWilliam}, A. 2009, \aap, 506, 729

\bibitem[{{Koch} {et~al.}(2012){Koch}, {L{\'e}pine}, \& {{\c C}al{\i}{\c
  s}kan}}]{Koch2012}
{Koch}, A., {L{\'e}pine}, S., \& {{\c C}al{\i}{\c s}kan}, {\c S}. 2012, in
  European Physical Journal Web of Conferences, Vol.~19, European Physical
  Journal Web of Conferences, 03002

\bibitem[{{Koch} {et~al.}(2016){Koch}, {McWilliam}, {Preston}, \&
  {Thompson}}]{Koch2016}
{Koch}, A., {McWilliam}, A., {Preston}, G.~W., \& {Thompson}, I.~B. 2016, \aap,
  587, A124


\bibitem[{{Kunder} {et~al.}(2012){Kunder}, {Koch}, {Rich}, {de Propris},
  {Howard}, {Stubbs}, {Johnson}, {Shen}, {Wang}, {Robin}, {Kormendy}, {Soto},
  {Frinchaboy}, {Reitzel}, {Zhao}, \& {Origlia}}]{Kunder2012}
{Kunder}, A., {Koch}, A., {Rich}, R.~M., {et~al.} 2012, \aj, 143, 57

\bibitem[{{Kunder} {et~al.}(2014{\natexlab{a}}){Kunder}, {Bono}, {Piffl},
  {Steinmetz}, {Grebel}, {Anguiano}, {Freeman}, {Kordopatis}, {Zwitter},
  {Scholz}, {Gibson}, {Bland-Hawthorn}, {Seabroke}, {Boeche}, {Siebert},
  {Wyse}, {Bienaym{\'e}}, {Navarro}, {Siviero}, {Minchev}, {Parker}, {Reid},
  {Gilmore}, {Munari}, \& {Helmi}}]{Kunder2014Tidal}
{Kunder}, A., {Bono}, G., {Piffl}, T., {et~al.} 2014{\natexlab{a}}, \aap, 572,
  A30

\bibitem[{{Kunder} {et~al.}(2014{\natexlab{b}}){Kunder}, {Rich}, {Gerhard},
  {Johnson}, {Chiappini}, {Martinez-Valpuesta}, {Martin}, {Ibata}, {Shen},
  {Li}, \& {De Propris}}]{Kunder2014Streams}
{Kunder}, A., {Rich}, R.~M., {Gerhard}, O., {et~al.} 2014{\natexlab{b}}, in
  Formation and Evolution of the Galactic Bulge, proceedings of a conference
  held 20-24 January, 2014 at the Sexten Center for Astrophysics, 15

\bibitem[{{Kunder} {et~al.}(2017){Kunder}, {Kordopatis}, {Steinmetz},
  {Zwitter}, {McMillan}, {Casagrande}, {Enke}, {Wojno}, {Valentini},
  {Chiappini}, {Matijevi{\v c}}, {Siviero}, {de Laverny}, {Recio-Blanco},
  {Bijaoui}, {Wyse}, {Binney}, {Grebel}, {Helmi}, {Jofre}, {Antoja}, {Gilmore},
  {Siebert}, {Famaey}, {Bienaym{\'e}}, {Gibson}, {Freeman}, {Navarro},
  {Munari}, {Seabroke}, {Anguiano}, {{\v Z}erjal}, {Minchev}, {Reid},
  {Bland-Hawthorn}, {Kos}, {Sharma}, {Watson}, {Parker}, {Scholz}, {Burton},
  {Cass}, {Hartley}, {Fiegert}, {Stupar}, {Ritter}, {Hawkins}, {Gerhard},
  {Chaplin}, {Davies}, {Elsworth}, {Lund}, {Miglio}, \& {Mosser}}]{Kunder2017}
{Kunder}, A., {Kordopatis}, G., {Steinmetz}, M., {et~al.} 2017, \aj, 153, 75

\bibitem[{{Kuzma} {et~al.}(2016){Kuzma}, {Da Costa}, {Mackey}, \&
  {Roderick}}]{Kuzma2016}
{Kuzma}, P.~B., {Da Costa}, G.~S., {Mackey}, A.~D., \& {Roderick}, T.~A. 2016,
  \mnras, 461, 3639

\bibitem[{{Lauberts} {et~al.}(1981){Lauberts}, {Holmberg}, {Schuster}, \&
  {West}}]{Lauberts1981}
{Lauberts}, A., {Holmberg}, E.~B., {Schuster}, H.-E., \& {West}, R.~M. 1981,
  \aaps, 43, 307

\bibitem[{{Mar{\'{\i}}n-Franch} {et~al.}(2009){Mar{\'{\i}}n-Franch},
  {Aparicio}, {Piotto}, {Rosenberg}, {Chaboyer}, {Sarajedini}, {Siegel},
  {Anderson}, {Bedin}, {Dotter}, {Hempel}, {King}, {Majewski}, {Milone},
  {Paust}, \& {Reid}}]{Marin-Franch2009}
{Mar{\'{\i}}n-Franch}, A., {Aparicio}, A., {Piotto}, G., {et~al.} 2009, \apj,
  694, 1498

\bibitem[{{Marino} {et~al.}(2008){Marino}, {Villanova}, {Piotto}, {Milone},
  {Momany}, {Bedin}, \& {Medling}}]{Marino2008}
{Marino}, A.~F., {Villanova}, S., {Piotto}, G., {et~al.} 2008, \aap, 490, 625

\bibitem[{{Massari} {et~al.}(2017){Massari}, {Mucciarelli}, {Dalessandro},
  {Bellazzini}, {Cassisi}, {Fiorentino}, {Ibata}, {Lardo}, \&
  {Salaris}}]{Massari2017}
{Massari}, D., {Mucciarelli}, A., {Dalessandro}, E., {et~al.} 2017, \mnras,
  468, 1249

\bibitem[{{Mauro} {et~al.}(2014){Mauro}, {Moni Bidin}, {Geisler}, {Saviane},
  {Da Costa}, {Gormaz-Matamala}, {Vasquez}, {Chen{\'e}}, {Cohen}, \&
  {Dias}}]{Mauro2014}
{Mauro}, F., {Moni Bidin}, C., {Geisler}, D., {et~al.} 2014, \aap, 563, A76

\bibitem[{{McWilliam}(2016)}]{McWilliam2016}
{McWilliam}, A. 2016, \pasa, 33, e040

\bibitem[{{Ness} {et~al.}(2013){Ness}, {Freeman}, {Athanassoula},
  {Wylie-de-Boer}, {Bland-Hawthorn}, {Asplund}, {Lewis}, {Yong}, {Lane}, \&
  {Kiss}}]{Ness2013}
{Ness}, M., {Freeman}, K., {Athanassoula}, E., {et~al.} 2013, \mnras, 430, 836

\bibitem[{{Nissen} \& {Schuster}(2010)}]{NissenSchuster2010}
{Nissen}, P.~E. \& {Schuster}, W.~J. 2010, \aap, 511, L10

\bibitem[{{Prusti} {et~al.}(2016){Prusti}, {de Bruijne}, {Brown}, {Vallenari},
  {Babusiaux}, {Bailer-Jones}, {Bastian}, {Biermann}, {Evans}, \&
  et~al.}]{GaiaDR1}
{Prusti}, T., {de Bruijne}, J.~H.~J., {Brown}, A.~G.~A., {et~al.} 2016, \aap,
  595, A1

\bibitem[{{Pryor} \& {Meylan}(1993)}]{Pryor1993}
{Pryor}, C. \& {Meylan}, G. 1993, in Astronomical Society of the Pacific
  Conference Series, Vol.~50, Structure and Dynamics of Globular Clusters, ed.
  S.~G. {Djorgovski} \& G.~{Meylan}, 357

\bibitem[{{Ruchti} {et~al.}(2010){Ruchti}, {Fulbright}, {Wyse}, {Gilmore},
  {Bienaym{\'e}}, {Binney}, {Bland-Hawthorn}, {Campbell}, {Freeman}, {Gibson},
  {Grebel}, {Helmi}, {Munari}, {Navarro}, {Parker}, {Reid}, {Seabroke},
  {Siebert}, {Siviero}, {Steinmetz}, {Watson}, {Williams}, \&
  {Zwitter}}]{Ruchti2010}
{Ruchti}, G.~R., {Fulbright}, J.~P., {Wyse}, R.~F.~G., {et~al.} 2010, \apjl,
  721, L92

\bibitem[{{Schlafly} \& {Finkbeiner}(2011)}]{Schlafly2011}
{Schlafly}, E.~F. \& {Finkbeiner}, D.~P. 2011, \apj, 737, 103

\bibitem[{{Schlegel} {et~al.}(1998){Schlegel}, {Finkbeiner}, \&
  {Davis}}]{Schlegel1998}
{Schlegel}, D.~J., {Finkbeiner}, D.~P., \& {Davis}, M. 1998, \apj, 500, 525

\bibitem[{{Sharples} {et~al.}(1990){Sharples}, {Walker}, \&
  {Cropper}}]{Sharples1990}
{Sharples}, R., {Walker}, A., \& {Cropper}, M. 1990, \mnras, 246, 54

\bibitem[{{Shetrone} {et~al.}(2009){Shetrone}, {Siegel}, {Cook}, \&
  {Bosler}}]{Shetrone2009}
{Shetrone}, M.~D., {Siegel}, M.~H., {Cook}, D.~O., \& {Bosler}, T. 2009, \aj,
  137, 62

\bibitem[{{Siebert} {et~al.}(2011){Siebert}, {Williams}, {Siviero}, {Reid},
  {Boeche}, {Steinmetz}, {Fulbright}, {Munari}, {Zwitter}, {Watson}, {Wyse},
  {de Jong}, {Enke}, {Anguiano}, {Burton}, {Cass}, {Fiegert}, {Hartley},
  {Ritter}, {Russel}, {Stupar}, {Bienaym{\'e}}, {Freeman}, {Gilmore}, {Grebel},
  {Helmi}, {Navarro}, {Binney}, {Bland-Hawthorn}, {Campbell}, {Famaey},
  {Gerhard}, {Gibson}, {Matijevi{\v c}}, {Parker}, {Seabroke}, {Sharma},
  {Smith}, \& {Wylie-de Boer}}]{Siebert2011}
{Siebert}, A., {Williams}, M.~E.~K., {Siviero}, A., {et~al.} 2011, \aj, 141,
  187

\bibitem[{{Sneden} {et~al.}(2014){Sneden}, {Lucatello}, {Ram}, {Brooke}, \&
  {Bernath}}]{Sneden2014}
{Sneden}, C., {Lucatello}, S., {Ram}, R.~S., {Brooke}, J.~S.~A., \& {Bernath},
  P. 2014, \apjs, 214, 26

\bibitem[{{Sneden}(1973)}]{Sneden1973}
{Sneden}, C.~A. 1973, PhD thesis, The University of Texas at Austin.

\bibitem[{{Timmes} {et~al.}(1995){Timmes}, {Woosley}, \& {Weaver}}]{Timmes1995}
{Timmes}, F.~X., {Woosley}, S.~E., \& {Weaver}, T.~A. 1995, \apjs, 98, 617

\bibitem[{{V{\'a}squez} {et~al.}(2015){V{\'a}squez}, {Zoccali}, {Hill},
  {Gonzalez}, {Saviane}, {Rejkuba}, \& {Battaglia}}]{Vasquez2015}
{V{\'a}squez}, S., {Zoccali}, M., {Hill}, V., {et~al.} 2015, \aap, 580, A121

\bibitem[{{Venn} {et~al.}(2004){Venn}, {Irwin}, {Shetrone}, {Tout}, {Hill}, \&
  {Tolstoy}}]{Venn2004}
{Venn}, K.~A., {Irwin}, M., {Shetrone}, M.~D., {et~al.} 2004, \aj, 128, 1177

\bibitem[{{Warren} \& {Cole}(2009)}]{Warren2009}
{Warren}, S.~R. \& {Cole}, A.~A. 2009, \mnras, 393, 272

\bibitem[{{Zacharias} {et~al.}(2017){Zacharias}, {Finch}, \&
  {Frouard}}]{Zacharias2017}
{Zacharias}, N., {Finch}, C., \& {Frouard}, J. 2017, \aj, 153, 166

\bibitem[{{Zoccali} {et~al.}(2017){Zoccali}, {Vasquez}, {Gonzalez}, {Valenti},
  {Rojas-Arriagada}, {Minniti}, {Rejkuba}, {Minniti}, {McWilliam}, {Babusiaux},
  {Hill}, \& {Renzini}}]{Zoccali2017}
{Zoccali}, M., {Vasquez}, S., {Gonzalez}, O.~A., {et~al.} 2017, \aap, 599, A12

\bibitem[{{Zwitter} {et~al.}(2004){Zwitter}, {Castelli}, \&
  {Munari}}]{Zwitter2004}
{Zwitter}, T., {Castelli}, F., \& {Munari}, U. 2004, \aap, 417, 1055

\end{thebibliography}
\end{document}